\documentclass[preprint,12pt]{elsarticle}

\usepackage{amssymb}

\journal{Physics A}

\begin{document}

\begin{frontmatter}

\title{Effective Mechanism for Social Recommendation of News}

\author[unifr]{Dong~Wei}
\author[unifr,uestc,ustc]{Tao~Zhou}
\author[unifr]{Giulio~Cimini\corref{cor1}}
\ead{Giulio.Cimini@unifr.ch}
\author[unifr]{Pei~Wu}
\author[unifr]{Weiping~Liu}
\author[unifr,uestc]{Yi-Cheng~Zhang}

\cortext[cor1]{Corresponding author}

\address[unifr]{Physics Department, University of Fribourg, CH-1700 Fribourg, Switzerland}
\address[uestc]{Web Sciences Center, University of Electronic Science and Technology of China, Chengdu 610054, P. R. China}
\address[ustc]{Department of Modern Physics, University of Science and Technology of China, Hefei 230026, P. R. China}

\begin{abstract}
Recommendation systems represent an important tool for news distribution on the Internet. In this work we modify 
a recently proposed social recommendation model in order to deal with no explicit ratings of users on news. 
The model consists of a network of users which continually adapts in order to achieve an efficient news traffic. 
To optimize network's topology we propose different stochastic algorithms that are scalable with respect to the network's 
size. Agent-based simulations reveal the features and the performance of these algorithms. 
To overcome the resultant drawbacks of each method we introduce two improved algorithms and show that they can optimize 
network's topology almost as fast and effectively as other not-scalable methods that make use of much more information.
\end{abstract}


\end{frontmatter}

\section{Introduction}

In a news distribution system with a tremendous number of users, such as digg.com and reddit.com, thousands of news 
are released every hour. When a user logs on the system without any support, he is drowned in the flood of information 
immediately~\cite{corso}. Recommender systems emerged to help users deal with the information overload and find out content 
matching their interests~\cite{burke,sarwar,resnick}. 
In traditional recommender systems relevant news are delivered to users according to their profiles and 
historical activities, with recommendation mechanism designed in a centralized way: data analysis and 
recommendation decision are made by the system for the whole network. In contrast, in social recommender systems 
users submit and forward news to others, as in a peer-to-peer network. The core idea of peer-to-peer is that 
all nodes in the system have the same function: each node can both gain service from other nodes, as a client, 
and provide service to others, as a server. 

In this work we present a modified version of a previously introduced news recommender system~\cite{medo}. 
This system is naturally implemented in a peer-to-peer environment\footnote{The peer-to-peer model presented here 
can be developed in one of two possible ways: as a centralized system (with all information stored in a central server), 
or in a real peer-to-peer system, where information about a node is actually stored in the peer node itself and each node 
both provides and gains services from other nodes.}: each user, considered as a node of the network, 
does not get news directly from some central servers, but he can post news and recommend them to other users 
(i.e. his followers) and at the same time he receives news from other users (i.e. his leaders). 
Hence each node acts as a follower and at the same time can become a leader in the network. 
When a user receives a news from his leader and likes it, he forwards it to his followers; 
then, if these followers like the news too, they forward it further to their followers, and so on. 
Hence news diffusion over the network resembles an epidemic spreading~\cite{psatorras,zhou1}. 
As in peer-to-peer systems, the leader-follower network is varying in topology: links between leaders and followers 
may be changed as the network evolves. The system continually refines the neighborhood of each user until connections 
between taste mates are established, so that news can spread quickly from the original source to the users who are 
most interested in it. Network evolution is therefore driven from an interplay between topology and dynamics~\cite{garlaschelli}.
The leader updating procedure is a crucial part of the whole recommender mechanism. 
An effective updating method should choose for each user a leader with very similar tastes at very low cost. 
In this work we compare several heuristic methods in detail. We discuss their efficiency, performance and scalability 
(which are important aspects for real systems), and we analyze the properties of the evolving network structure 
with agent-based simulations. A comparison with other widely adopted popularity-based recommendation methods has already been discussed in~\cite{medo}.

This work is organized as follows. In Section 2 we present related work. 
In Section 3 we describe the news recommendation method and an agent-based model to test it. 
In Section 4 we report simulation results for network's evolution with different updating methods. 
In Section 5 we introduce two refined updating methods. Lastly, we conclude with discussion.

\section{Related Works}

So far, plenty of recommendation techniques have been proposed by researchers. 
These techniques can be classified in three main categories: content-based, collaborative filtering and social filtering methods. 

Content-based methods~\cite{pon,cantador} try to classify items according to their content (e.g. by keywords) and to extract users' preferences from their pick history. 
These two elements are used to obtain recommendation: items recommend to a user are the ones whose characteristics match user's preferences.

Collaborative filtering methods make predictions about the interests of a user by collecting taste information 
from the community the user belongs to, with the underlying assumption that those users who agreed in the past 
tend to agree again in the future. In memory-based collaborative filtering, such as in Amazon.com~\cite{linden}, 
the recommendation score for an item is computed as a weighted average of ratings given by other users with weights 
proportional to the similarity between users~\cite{resnick}. In model-based collaborative filtering methods 
(e.g. Bayesian clustering~\cite{breese} and PLSI~\cite{hofmann}) users are classified according to their preferences. 
Diffusion-based methods, where heat conduction~\cite{zhang}, random walk~\cite{zhou2} or hybrid processes~\cite{zhou3} are employed 
to efficiently extract the taste similarity between users or items, can be considered as extensions of collaborative filtering. 

Social filtering methods make use of users' relationship in a social network to obtain recommendations. 
These methods are based on the assumption that users prefer recommendations made by a friend than those by a centralized 
system~\cite{sinha}. In other words, social influence is more determinant for users to decide what to read than similarity 
of past activities~\cite{borgatti,gulli}. Generally, a social recommender system can be devised in two different ways: pull and push. 
In the first case, each user selects other users as his information sources and he can pull information 
from these sources (as in, e.g., Delicious.com). In the latter, each user can push what he likes to other users 
who have accepted him as an information source (as in, e.g., Twitter.com). In many of these systems users have also the possibility 
to assign information with user-defined words, so-called \emph{tags}, that allow a more effective management of resources by users~\cite{capocci,zhangZ1,zhangZ2}.

Different from all methods mentioned above, the news recommendation mechanism we study in this work combines 
memory-based and social recommendation together with network's topology optimization. 
Recommendation scores are computed using  users' past ratings, though recommendations come only from the information sources 
a user selects; in parallel, the leader-follower network is continually refined according to similarity of user tastes, 
in order for news to be efficiently delivered to their eagerest readers.
This combination of social and similarity recommendation turns out to be particularly effective: the presence of directed links among users 
makes the recommendation process naturally fast and scalable, while the use of rating patterns' similarity in the information diffusion process 
enhance users' social contacts and experience.

\section{Model Description}

In this section we describe the news recommendation mechanism~\cite{medo} that provides a personalized set of news 
for each user. As stated in the introduction, the system consists of a set of users with distinct tastes 
who continually post and evaluate various news items. 

\subsection{Similarity Estimation}

The taste similarity between users is predicted using statistical correlation of users' rating histories. 
We adopt Pearson correlation coefficient: for a pair of users $i$ and $j$ 
\begin{equation}
\label{eq1}
s_{i,j}=\frac{\sum_c(r_{i,c}-\overline{r}_i)(r_{j,c}-\overline{r}_j)}
{\sqrt{\sum_c(r_{i,c}-\overline{r}_i)^2}\sqrt{\sum_c(r_{j,c}-\overline{r}_j)^2}},
\end{equation}
where $r_{i,c}$ and $r_{j,c}$ are the ratings given by these users to news $c$ and 
$\overline{r}_i$ and $\overline{r}_j$ are the average ratings of $i$ and $j$. 
$s_{i,j}\in[-1,1]$, with $s_{i,j}>0$ indicating that user $i$ and $j$ tend to agree in evaluations of news 
while with $s_{i,j}<0$ that they tend to disagree. Despite the network is directed, $s_{i,j}=s_{j,i}$. 

The similarity score (\ref{eq1}) is different from the one used in~\cite{medo}. 
This is because we assume that users evaluate news in a binary scale: $r_{i,c}=1$ if user $i$ likes news $c$ 
and otherwise $r_{i,c}=0$. In other words, we do not distinguish between a disliked and not rated news since 
in many real news distributing systems the only available information about users' ratings is users' \emph{click history}. 
Therefore the simples possible assumption is to treat a user's click on a news as a positive vote 
for the news itself, while any information about users' negative interests remain undiscovered. 

The choice of Pearson correlation (\ref{eq1}) as similarity metric is motivated by the fact that it can be used 
also in systems where users gives explicit ratings to news. However the choice of other metrics, such as 
Jaccard coefficient or Cosine similarity, does not alter the behavior and the features of the system. 
This statements also holds for asymmetric similarity metrics (as for example the asymmetric Jaccard coefficient), 
because the difference of rating patterns between similar users becomes smaller and smaller as they evaluate more news. 
This causes the asymmetric $s_{i,j}$ and $s_{j,i}$ to converge to the same value, i.e. to a symmetric coefficient. 

We use some heuristics methods to avoid problems related to data sparsity. 
If two users have no evaluations $s_{i,j}$ is undefined. 
Therefore at the beginning of system's evolution we set $s_{i,j}=s_0$, with $s_0$ a small fixed value, for each pair of users. 
$s_{i,j}$ is undefined also when a user gives all positive or negative evaluations. In this case we set $s_{i,j}=-1$ 
to prevent this kind of malicious behavior (i.e. giving ratings that do not bring any information about the user's 
interests) from affecting recommender system's performance.

Note that for the sake of system scalability we do not store the similarity scores between all user pairs 
but only the reading history of users, from which the similarity scores are computed. 
In other words, the information is distributed over the network: the click history of each node, 
which is stored in the peer node itself, is available to other nodes and can be retrieved by them in order 
to perform local tasks like assigning recommendation scores to incoming news or updating connections to other nodes. 

\subsection{News Recommendations and Network Updating}

When a user logs into the recommender system, the system checks all the candidate unread news forwarded by his leaders in his 
waiting list and then selects the top $k$~\cite{blum} of them to recommend him according to their recommendation scores. 
The recommendation score of each news is equal to the similarity between the user and the leader who forwarded it, 
so that a high recommendation score is assigned to the news sent by leaders with high similarity. 
When the same news is recommended to a user from multiple sources, recommendation scores 
are not summed up, as in~\cite{medo}, to prevent Eclipse attacks on the system~\footnote{In an Eclipse attack 
a modest number of malicious nodes conspire to fool correct nodes into adopting the malicious nodes as their peers, with the goal of 
dominating the neighbor sets of all correct nodes.}~\cite{singh}. Instead, each news can be recommended only once to each user. 
When the user likes a recommended news, he will forward it to the waiting list of his followers. 
Also, he can introduce a new news item in the system and forward it to his followers. 
We employ a book-keeping mechanism to make sure that each user does not receive news he has already read. 
Besides, users' waiting lists are of limited length, and when they are completely filled with candidate news 
the later arrived (i.e. the fresh news) have few chances to jump on top of the recommender list and be consequently read. 
To allow fresh contents to be more visible, the waiting list of a user is cleaned up after the user gets offline. 
Other mechanisms to promote the diffusion of novel news items are discussed in~\cite{medo}.

As news spread over the network and get evaluated, taste differences between users and their leaders can be gradually 
identified. We use this information to refine the local neighborhood network and improve news propagation: 
from time to time we find the least similar leaders of all users and replace them with new ones. 
In what follows we will compare several updating methods in detail.

We remark that the present model differs from the one described in~\cite{medo} under several aspects, 
as we make use of much less information, for instance: ratings are binary, the different recommendation scores for a news are not aggregated, 
recommendation list of users are cleaned every time they get offline. By keeping only the necessary information and using fewer computational resources, 
we want our system to be efficient and fast as well as effective and robust. 

\subsection{Agent-based Modeling}

We make use of an agent-based approach to test the model. In order to depict the tastes of users and the attributes of news, 
we use $D$-dimensional vectors containing only 0 or 1. User $i$'s taste vector ($\vec{t}_i$) describes what topics 
he likes~(1) or dislikes~(0). The maximum number of distinct taste vectors is $2^D$. In~\cite{medo} an alternative setting 
with a fixed number of 1s in each vector was used. However the present scenario better depicts users who are heterogeneous 
in their scope of interests. 

At the beginning of the simulation, $L=D$ randomly selected leaders are assigned to each user. 
The similarity between each user and his newly selected leaders is set to the uniform default value $s_0$. 
At every running step each user gets online with probability $p_o$ and, if active, submits a new news item with probability 
$p_s$. The attribute vector of the submitted news is identical to the taste vector of the user who submitted it. 
When a user is online, he reads and evaluates the top $k$ news recommended by his leaders. 
The overlap between user's tastes and news' attributes, computed by counting the number of identical bits 
in the corresponding vectors, is used to obtain user's evaluation on the news item. 
Agent $i$ likes news $c$ with attribute vector $\vec{a}_c$ if: 
\begin{equation}
\label{eq2}
\|\vec{t}_i-\vec{a}_c\|\leq \theta
\end{equation}
here, $\|\cdot\|$ is the norm of the vector and $\theta$ is the approval threshold. 
If agent $i$ likes news $c$, then $r_{ic}=1$; if $i$ dislikes $c$ or if $i$ has not evaluated $c$ yet, $r_{ic}=0$. 
If a user likes a news, he forwards it to the waiting lists of his followers. 
Users' assessments on news are used to compute the similarity scores between users and their leaders. 
The leader updating procedure takes place for each user after this user gets online $u$ times. 

\subsection{Performance Metrics}

For any recommender system, the approval fraction, i.e. how often users are satisfied with the news they get recommended, 
is the most important performance measure. Besides, we focus mainly on the properties of the leader-follower network. 
Since in a computer simulation we have the luxury of knowing the taste vectors of users, we can compute the average value 
of the taste vector differences between a user and his leaders. Since there are no identical users in the system, 
the baseline value of these differences is 1 and we define the excess differences as the average number of differences 
minus~1. This measure shows the intrinsic quality of the leadership in the network: the lower the average differences, 
the higher the taste likeness between leaders and their followers. 

\section{Simulation Analysis}

In our simulations we assume 13-dimensional taste vectors and hence the system consists of 8192 users. 
For other simulation parameters, see table \ref{tab1}. 

\begin{table}[h!]
\caption{Simulation parameters and their default values}\label{tab1}
\centering
\begin{tabular}{l l l}
\hline\hline
Parameter						&	Notation	&	Value	\\
\hline\hline
Network size						&	$N=2^D$		&	$8192$	\\
Length of users' taste / news' attribute vector		&	$D$		&	$13$	\\
Number of leaders for each user				&	$L$		&	$13$	\\
Online probability at each step				&	$p_o$		&	$0.02$	\\
Submitting probability when online			&	$p_s$		&	$0.01$	\\
Period of leader updating				&	$u$		&	$10$	\\
Length of waiting lists					&	$l$		&	$100$	\\
News read by users when online				&	$k$		&	$10$	\\
Acceptance threshold					&	$\theta$	&	$6.5$	\\
Initial similarity value				&	$s_0$		&	$0.1$	\\
\hline
\end{tabular}
\end{table}

\subsection{Leader Updating Methods}

Updating the leader-follower network consists in finding for a specific user $i$ the least similar leader (e.g. user $j$) 
and replace this leader with a new one (e.g. user $k$). Some straightforward methods to select the new leader are:
\begin{itemize}
  \item \emph{Pure Random}: the new leader is simply a randomly selected user; 
  \item \emph{Selective Random}: the new leader is selected as in the Pure Random method, but he replaces the old leader 
  only if he has a higher similarity score with user $i$, i.e. if $s_{ik}>s_{ij}$;
  \item \emph{Best's Random Leader (BRL)}: the new leader $k$ is a user chosen at random 
  among the leaders of user $i$'s best leader;
  \item \emph{Global Memory Based (GMB)}\footnote{GMB updating method is similar to a top $L$~\cite{blum} Collaborative Filtering algorithm 
(although in this case there is no spreading of news on a social network).}: the new leader $k$ is the user in the network with the highest similarity 
  with user $i$. 
\end{itemize}
 
\begin{figure}[t!]
\centering
\includegraphics[width=300pt]{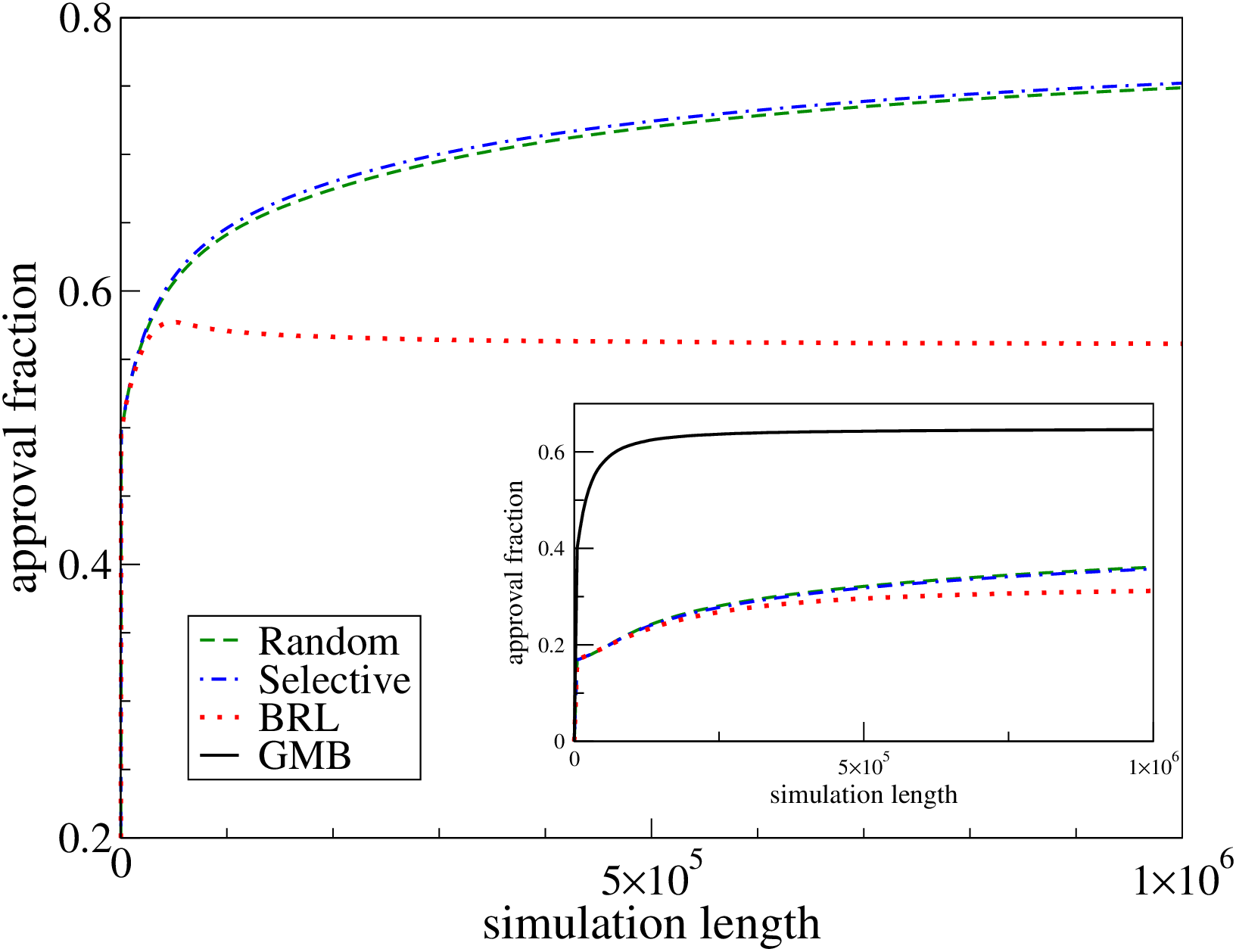}
\includegraphics[width=300pt]{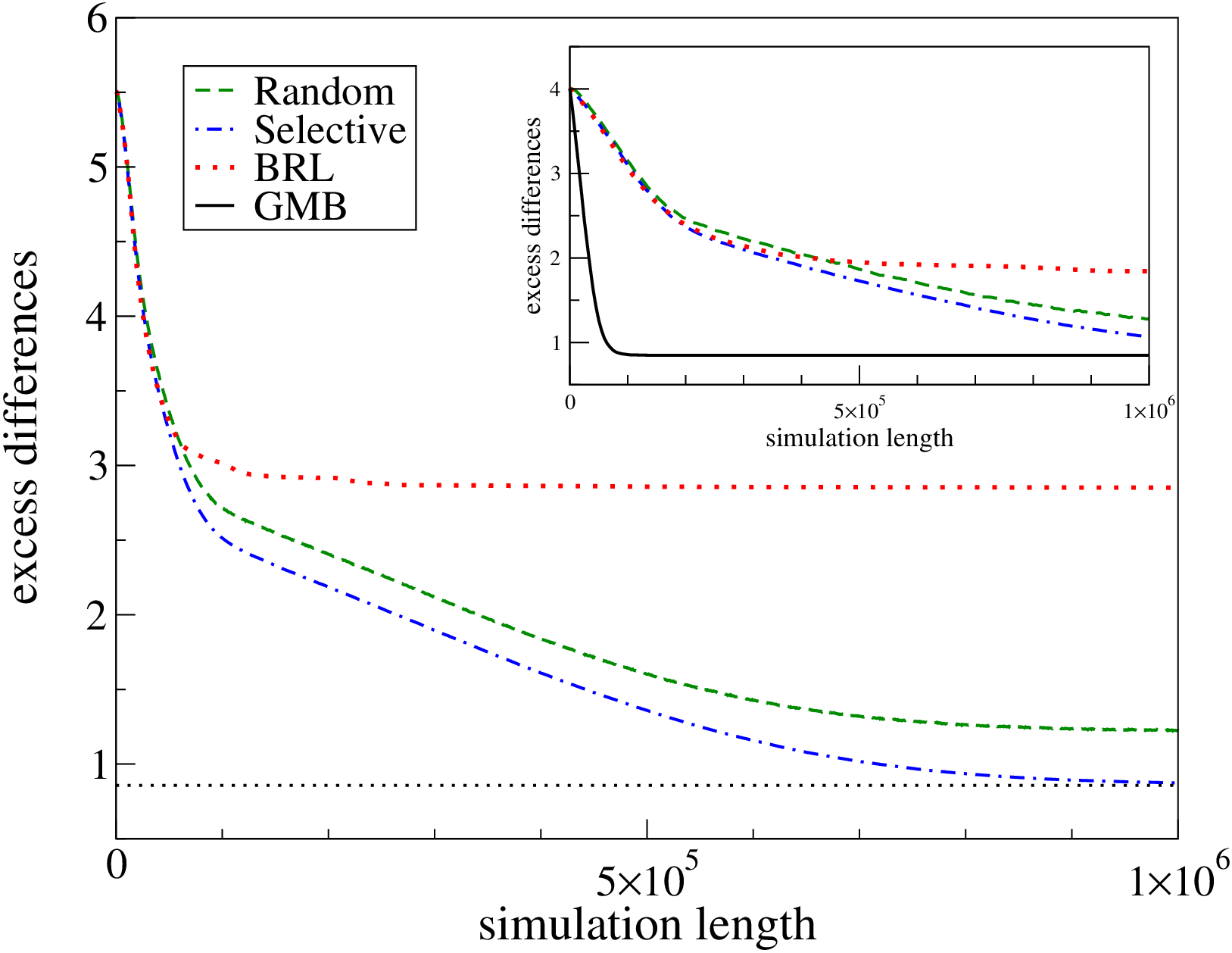}
\caption{Approval fraction and excess differences in the network for different leader updating methods. 
The horizontal dotted line represents the optimal assignment of leaders. 
Insets: comparison of these methods with the GMB for a smaller system ($D=10$).}\label{fig1}
\end{figure}

Figure \ref{fig1} shows the performance of these methods. 
As expected, the GMB strategy performs best (by achieving the highest approval fraction and the best assignment of 
leaders---see the insets of figure \ref{fig1}) because it makes use of all available information. However, 
this method is very expensive in terms of computing cycles: it involves the computation of $N-1$ similarity scores 
to update the leaders of a single user, and $O(N^2)$ to update the whole network once. Therefore it is not scalable 
with respect to the system's size. We use it only as a benchmark for the other methods. The pure and selective random 
methods perform worse, but have a much smaller computational cost---$O(N)$ to update the whole network---and can still 
converge to a network state with a good assignment of leaders. As compared to pure random, the selective method may be 
slower in network exploration but is more careful in selections, therefore it eventually reaches a network state as good as 
with the GMB method. On the other hand, the BRL method, while exploiting the local neighborhood network with computational 
cost $O(N*D)$ (it involves the computation of $D$ similarity values to identify the best leader of a user), reaches 
a strongly sub-optimal network state and has the worst performance in approval fraction. This is because 
with this procedure a user can only exchange his leader for a two-step away user, hence he gets a very limited range 
of possibilities for selection and may not ever get in contact with his taste mates if they are too distant from him. 

\begin{figure}[b!]
\centering
\includegraphics[width=250pt]{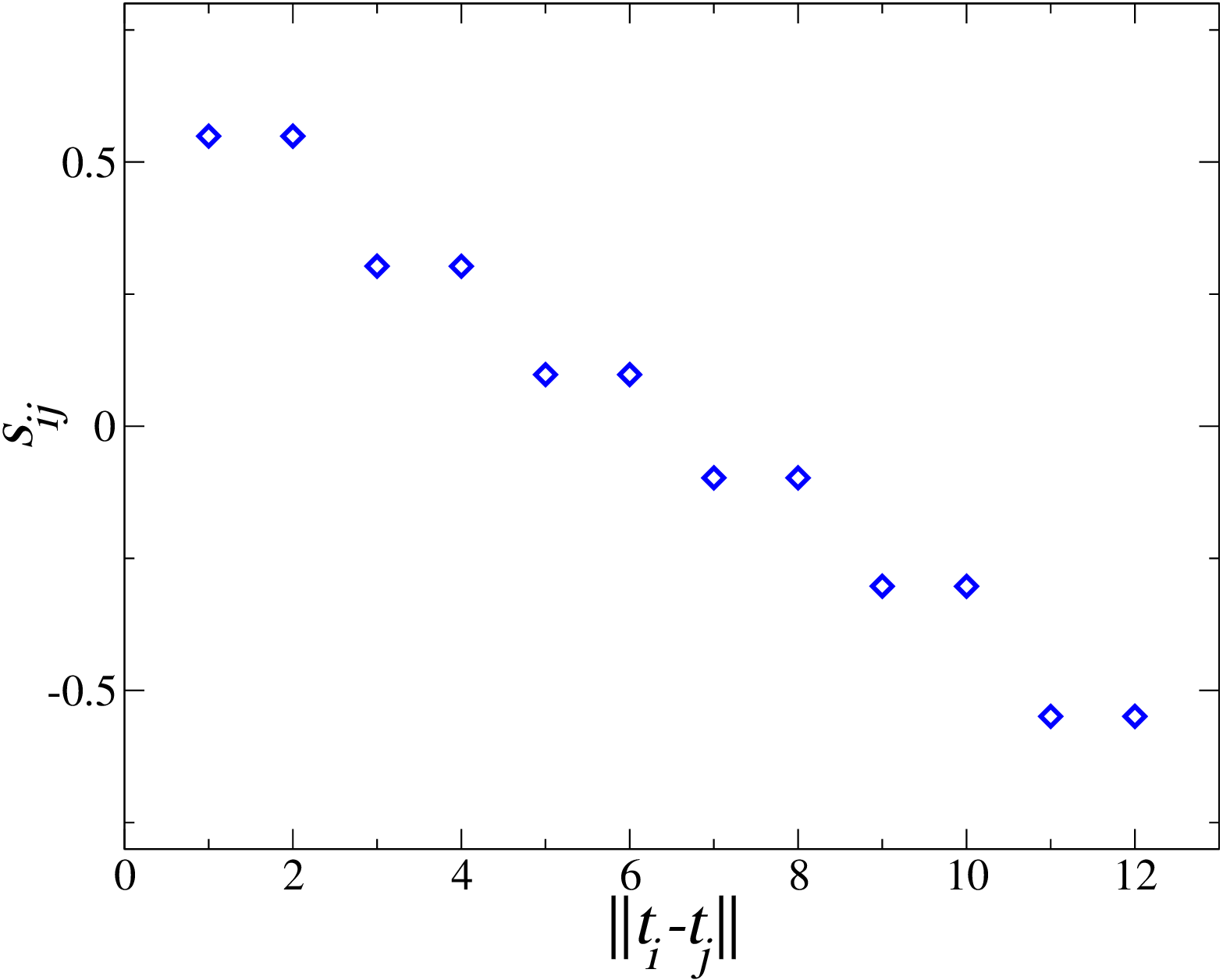}
\caption{Average similarity score (\ref{eq1}) versus vector differences averaged over all pairs of users. 
Similarities were computed over 1000 randomly selected news evaluated in common by all users.}\label{fig3}
\vspace{10pt}
\includegraphics[width=250pt]{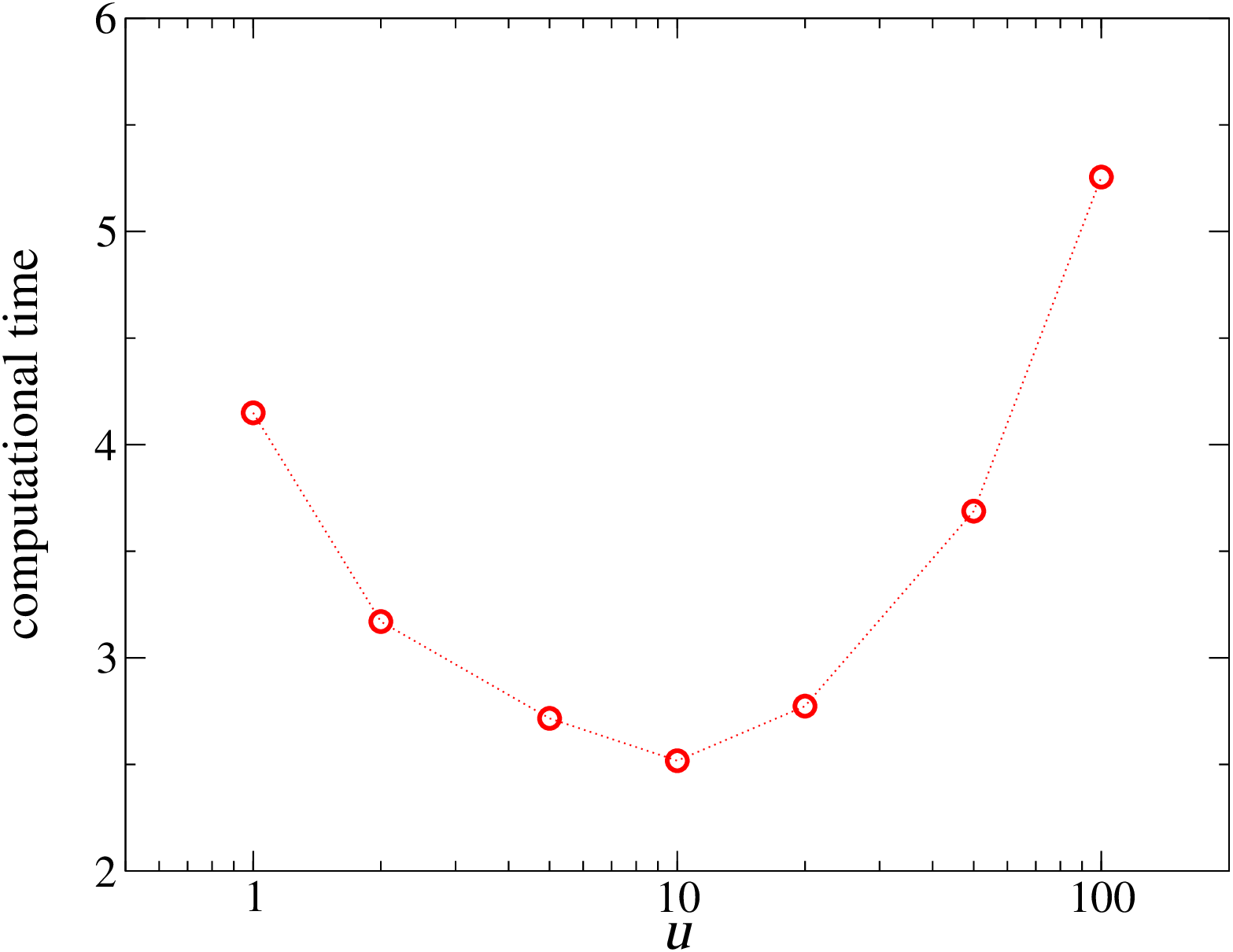}
\caption{Computational time (in arbitrary units) required to reach excess differences equal to 3 
for different values of the updating interval $u$. Simulations with the selective random updating method.}\label{fig4}
\end{figure}

Lower panel of figure \ref{fig1} also shows that system's evolution---driven by any leader updating method---eventually stops when an \emph{equilibrium state} is found. 
In such state, users have no advantage in changing their choices (i.e leaders) simply because they cannot find better ones. 
Thus an equilibrium state corresponds to an optimum for each agent.
However, since users make decisions independently from what others have chosen, 
the particular equilibrium state reached by the system does depend strongly on the updating method employed (i.e. how users make their decisions) 
and only weakly on the interaction between agents (as user $i$'s selection of leaders little influences news arriving to user $j$). 
Hence, if the updating method allows users to make the best choices for themselves, the equilibrium state will be not only an optimum for each agent but also a global minimum, 
i.e. a \emph{ground state} for the whole system. 
In the ground state each user is linked to his best taste mates: there is not a better possible configuration in terms of leader assignment to users. 
The excess differences in the ground state (represented in figure \ref{fig1} by a horizontal dotted line) are determined by equation (\ref{eq2}) 
in our artificial environment: as shown in figure~\ref{fig3}, for each user first-order and second-order taste mates 
(i.e. users who differ from him in one and two elements of taste vectors, respectively) are indistinguishable in terms of the similarity (\ref{eq1}). 
Since each user has $D$ first-order and $D(D-1)/2$ second-order taste mates 
in the network and he has to chose $D$ of them as leaders, the maximization of similarity is equally likely to select 
a first or second order taste mate and hence the expected value of excess differences is
\begin{equation}
\label{eq5}
1\cdot\frac{D}{D(D+1)/2}+2\cdot\frac{D(D-1)/2}{D(D+1)/2}-1=\frac{D-1}{D+1}
\end{equation}

Finally we remark that the frequency of the leader updating strongly influences the convergence rate of the system. 
If the updating interval is too short, there are not enough evaluations to assess the real similarity between a leader 
and his followers, hence the leadership structure of the network deteriorates. On the other hand, if the updating interval 
is too long there is redundant information for similarity evaluations and leader updating, therefore the evolution 
of the system is unnecessarily slowed down. Figure \ref{fig4} shows that it is possible to find an optimal value 
of the leader updating frequency which yields the best convergence rate.

\subsection{Network's Properties}

We now compare the leader updating methods from the viewpoint of network topology. 
We use two quantities to describe the network:
\begin{itemize}
 \item \emph{Link reciprocity}, the tendency of node pairs to form connections between each other~\cite{wasser}. 
This quantity is defined as the ratio between the number of bidirected links and the total number of links in a network:
\begin{equation}
\label{eq3b}
\rho=\frac{\sum_{i\neq j}\:a_{ij}\:a_{ji}}{\sum_{i\neq j}\:a_{ij}}
\end{equation}
where $a_{ij}=1$ if user $j$ is one of user $i$'s leaders ($a_{ij}=0$ otherwise). 
 \item \emph{Clustering coefficient}, the tendency of a network to form tightly connected neighborhoods~\cite{fagiolo}. 
This quantity is calculated as the ratio between the number of triangles a user forms with his neighbors 
and the total number of possible triangles this user can form, averaged over all users. 
When edges are directed, each user can generate up to eight different triangles with any pair of neighbors, hence:
\begin{equation}
\label{eq3a}
c=\frac{1}{N}\frac{\sum_{i\neq j\neq k}(a_{ij}+a_{ji})(a_{ik}+a_{ki})(a_{jk}+a_{kj})}{2[d_i^{tot}(d_i^{tot}-1)-2d_i^\leftrightarrow]}
\end{equation}
where $d_i^{tot}=\sum_j(a_{ij}+a_{ji})$ denotes the number of $i$'s neighbors ($i$'s total degree) and 
$d_i^\leftrightarrow=\sum_ja_{ij}a_{ji}$ is the number of bidirected links $i$ forms with his neighbors.
\end{itemize}

\begin{figure}[t!]
\centering
\includegraphics[width=300pt]{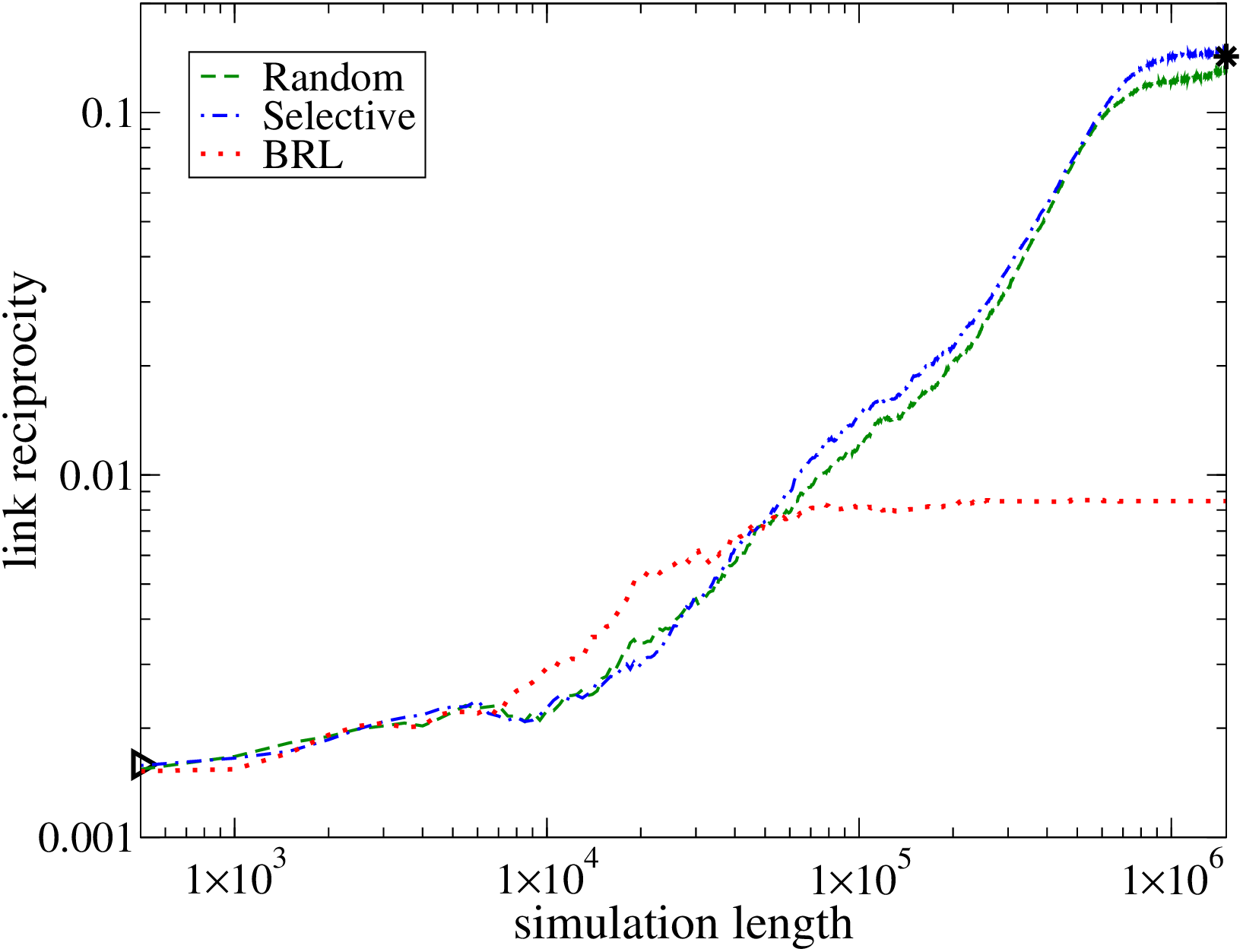}
\includegraphics[width=300pt]{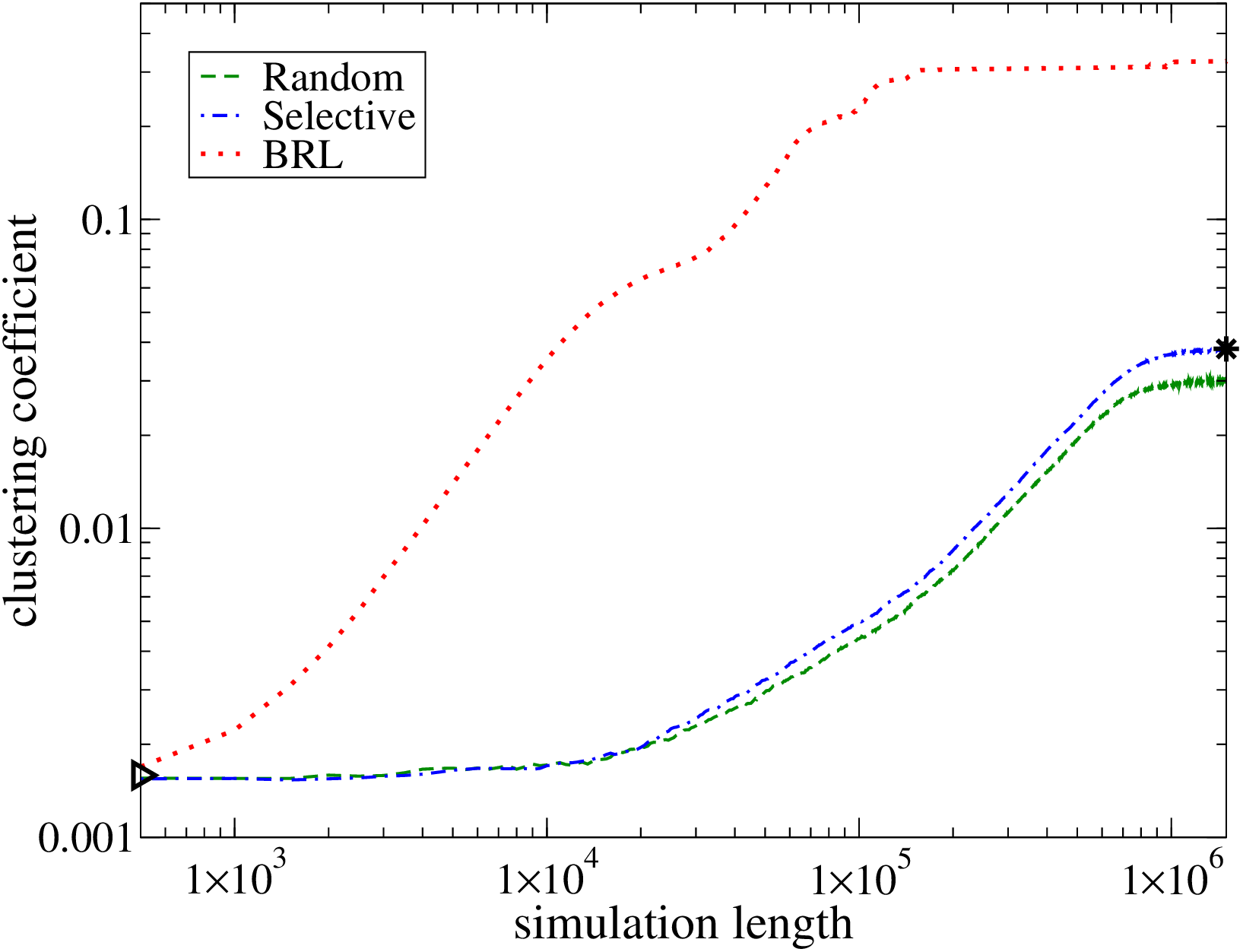}
\caption{Link reciprocity and clustering coefficient of the network for different leader updating methods ($\overline{a}$ 
is represented by $\triangleright$ and $\rho_{gs}$, $c_{gs}$ by $\ast$ in the respective panels).}\label{fig5}
\end{figure}

Figure \ref{fig5} shows the evolution of the link reciprocity in networks with different leader updating methods. 
In the initial random networks the average probability of finding a reciprocal link between two connected vertices 
is simply equal to the average probability of finding a link between any two vertices, which is given by 
$\overline{a}=\sum_{i \neq j}\:a_{ij}/[N(N-1)]=D/(N-1)$. Hence the starting value of $\rho$ is equal to $\overline{a}$.
As the network evolves, the reciprocity coefficient grows from $\overline{a}$ with all updating methods: 
network's structure becomes more and more reciprocal. This growth eventually stops when a stationary 
state of the network is found. If the network is driven to the ground state (e.g. by the selective method) 
the value of the reciprocity becomes $\rho_{gs}=L/[D(D+1)/2]=2/(D+1)$, because in this state each user choses 
his $L$ leaders among $D(D+1)/2$ best taste-mates who are indistinguishable in term of the similarity score 
(see figure \ref{fig3}): each user choose randomly among them and $\rho\rightarrow\rho_{gs}$.

The evolution of the clustering coefficient is also shown in figure \ref{fig5}. 
With any updating method $c$ grows from its initial value, again equal to $\overline{a}$ (the probability to find a link between two vertices in the initial random network). 
The selective method, driving the system to the optimal state, makes $c$ converge to $c_{gs}=\rho_{gs}P(\|\vec{t}_j-\vec{t}_k\|\leq 2\;|\;a_{ij}=a_{ik}=1)=8/[(D+1)(D+2)]$ 
(i.e. the probability to find a link between two users $j$ and $k$ who are linked to user $i$ and who are taste-mates of user $i$). 
$c_{gs}$, like $\rho_{gs}$, corresponds to users randomly choosing leaders among their best taste-mates. 
Instead with the BRL method, where links can only form locally, the clustering coefficient becomes immediately very large 
(even bigger than $c_{gs}$): network's evolution is hindered because a high clustering coefficient means that news 
do not spread effectively (they get few directions to propagate and their traffic is limited to a subset of the network). 
Users who happen to be far apart are unlikely to read the same news: 
their similarity remains undiscovered and they cannot be connected even if they are taste-mates. 
Note also that with this method once a node has no followers, it will never get any follower in future 
because it cannot be selected as a leader, hence some users may get isolated from the system. 
These two factors cause the network evolution to end in a sub-optimal equilibrium state.

\section{Refined Algorithms}

In this section we propose new updating methods that strongly enhance the system's performance. 

\subsection{Hybrid algorithm}

The BRL method has the ability to exploit the local neighborhood network, that is the most significant feature 
of a peer-to-peer system. Unfortunately, as we have mentioned, it has some major drawbacks caused by 
the limited range of the network that this method is able to explore. 
We attempt to overcome these problems by broadening the exploration range. The first possible solution 
is to explore the whole second-order neighborhood (i.e., not only the subset related to the best leader), 
or even to go further by including the third-order neighborhood, fourth-order neighborhood, and so on. 
However, the computational cost of such exploration grows very fast---it is $O(N*D^k)$ for a $k^{th}$-order 
neighborhood---and soon becomes comparable with the cost of the GMB method.
\begin{figure}[t!]
\centering
\includegraphics[width=300pt]{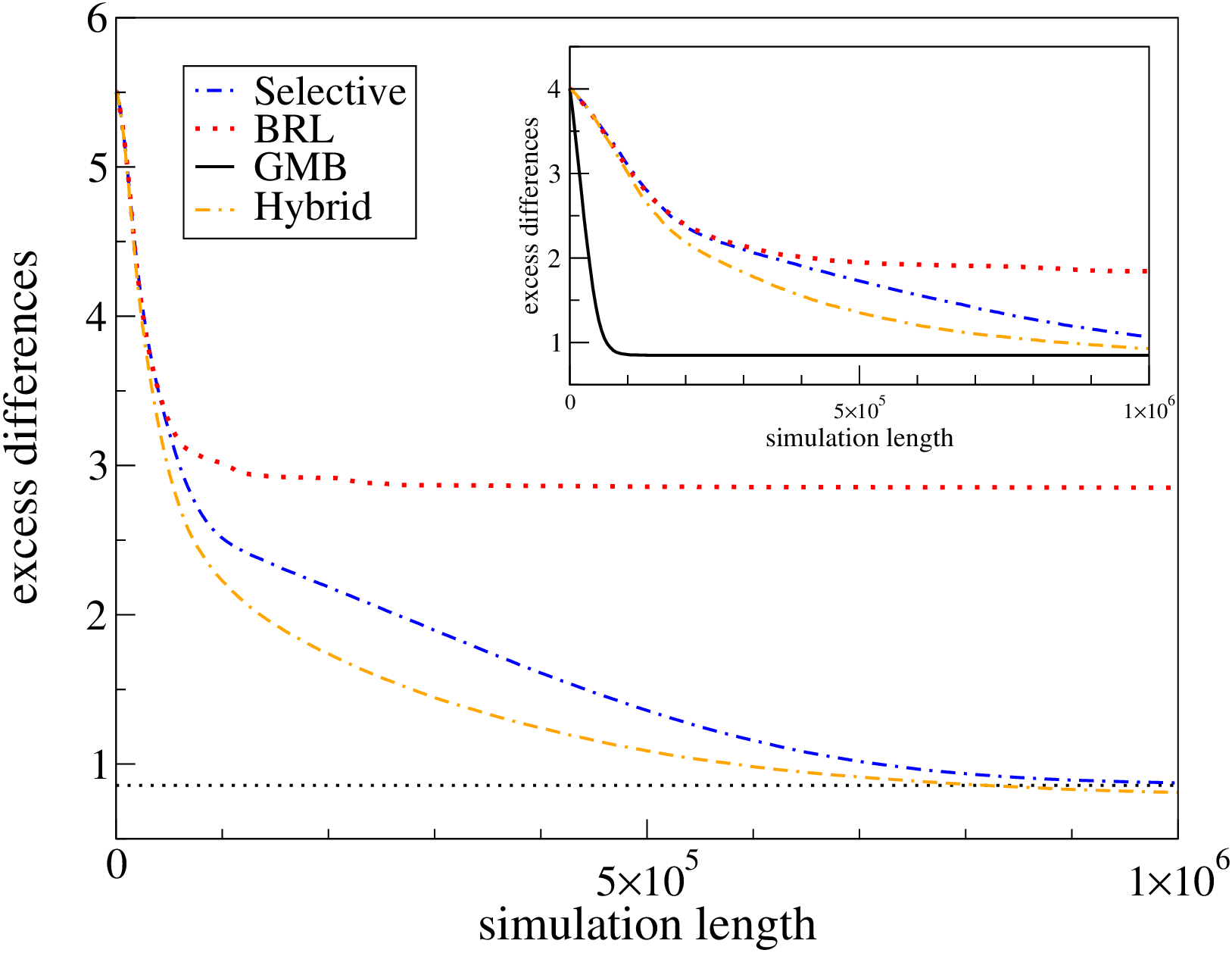}
\caption{Performance of the Hybrid algorithm ($p=0.75$) compared to other leader updating methods. 
Inset: comparison with GMB for a smaller system ($D=10$).}\label{fig7}
\end{figure}

A more effective solution is to employ some randomness in the updating method in order to have the possibility 
to connect users regardless of their distance and prevent the system from getting trapped in a sub-optimal state. 
Hence we propose a \emph{hybrid} updating method that combines selective and BRL methods: 
at each updating step we employ the first with probability $p$, otherwise we employ the second. 
This mechanism mimics the evolution of self-organizing communities where users search for friends 
among friends of friends but casual encounters also occur. 
Figure \ref{fig7} shows that the hybrid method improves system's performance as compared to both selective and BRL methods, 
still keeping an affordable $O(N*D)$ computational cost. Hence it represents a cheap and effective updating algorithm 
suitable for large systems.

\subsection{Las Vegas algorithm}

We now concentrate on the drawbacks of the above mentioned random updating methods. 
The pure random method may replace an old leader with a less similar one, hence it cannot reach the best 
assignment of leaders. Therefore the selective method outperforms the pure random. However the selective method 
may leave the network unchanged if it doesn't find a better leader (which may occur quite often), hence 
slowing down the system's evolution. A simple solution to this problem is to employ a selective approach with more trials. 
We name this new method \emph{Las Vegas algorithm}~\cite{babai}. 
Its basic idea is that if the newly randomly selected candidate leader is less similar than the old leader, 
we drop it and retry the random updating until a better candidate is found. 
Since there may not be a better candidate, we set the maximum number of trials to a predefined value $v$. 
The bigger this value, the longer the execution time, but also the higher chances to find a better leader.
\begin{figure}[b!]
\centering
\includegraphics[width=300pt]{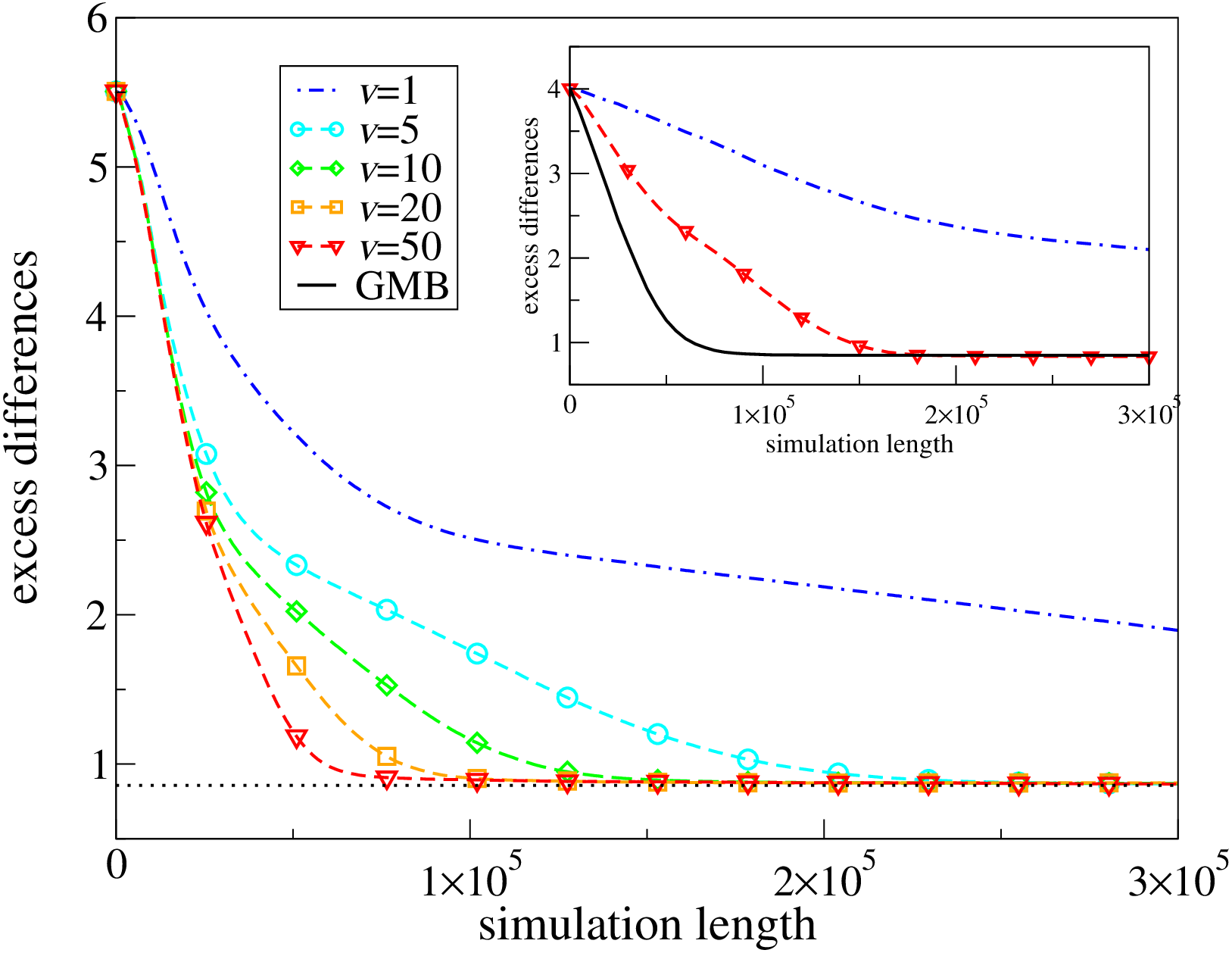}
\caption{Performance of the Las Vegas algorithm for different values of $v$ compared to other updating methods. 
Inset: comparison with GMB for a smaller system ($D=10$).}
\label{fig8}
\end{figure}

\begin{figure}[t!]
\centering
\includegraphics[width=240pt]{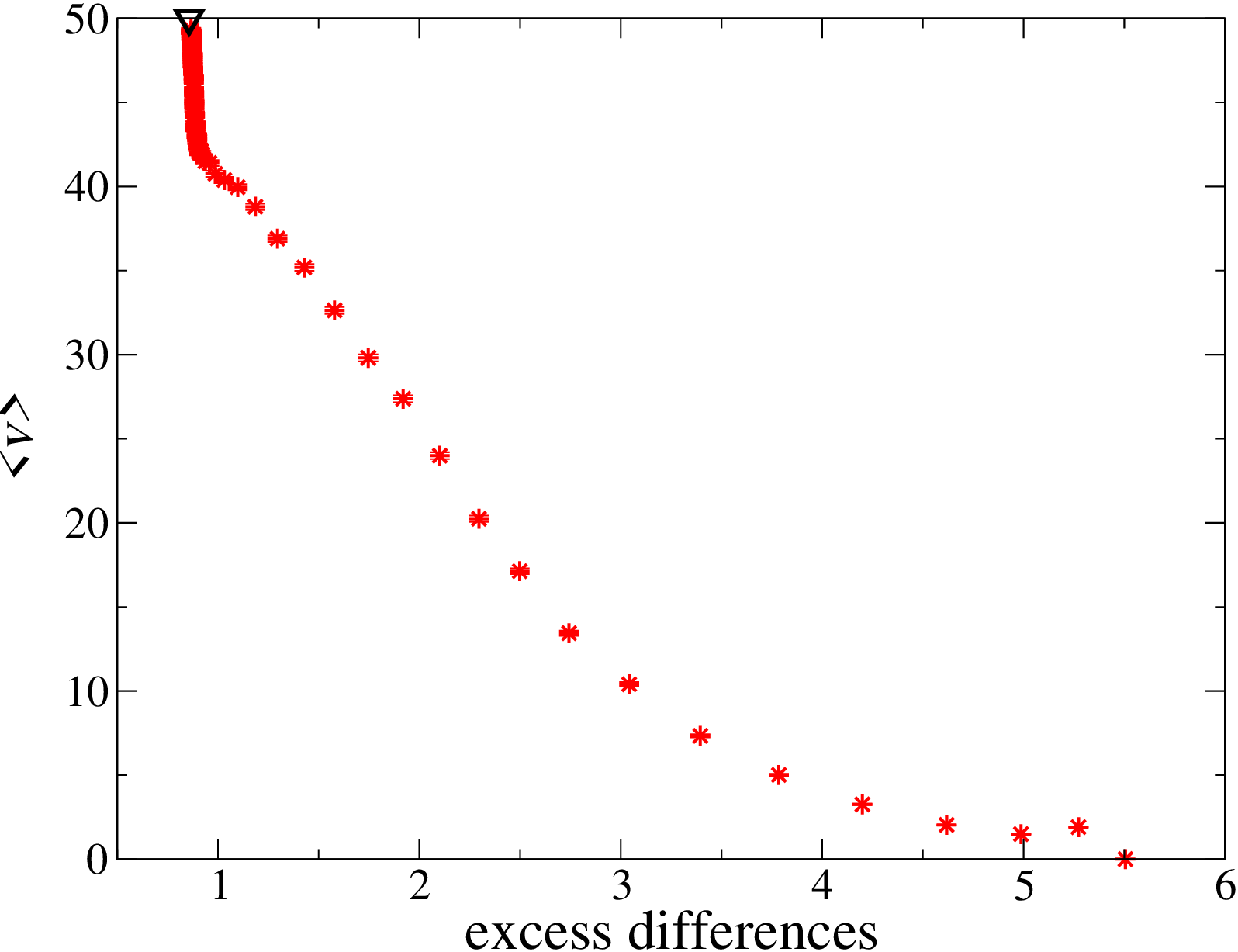}
\caption{Average number of actual trials $\langle v\rangle$ of the Las Vegas algorithm with $v=50$ at different stages 
of the system's evolution (represented by the value of the excess differences). 
At the beginning, $\langle v\rangle\ll v$ (the fictitious starting point at $\langle v\rangle=0$ only represent 
the initial network's state). As the system approaches the ground state, $\langle v\rangle\rightarrow v$ 
($\triangledown$ represents the point of convergence).}\label{fig9}
\end{figure}

Figure \ref{fig8} shows the performance of the Las Vegas algorithm for different values of $v$. 
Note that the selective method is nothing but a Las Vegas method with $v=1$. As $v$ increases from 1, 
the convergence rate of the Las Vegas method becomes faster and faster, and it becomes even comparable with the GMB method, 
yet keeping the affordable computational cost $O(N*v)\ll O(N^2)$ of GMB. 
Moreover, the average number of actual trials $\langle v\rangle$ of the Las Vegas algorithm (figure \ref{fig9}) 
is very small at the beginning of system's evolution, when it is very easy to find better candidate leaders for users: 
$\langle v\rangle\ll v$ again reduces computational cost. Then the network approaches to the ground state and 
$\langle v\rangle\rightarrow v$ as each user is already linked to his taste mates; 
however at this stage the evolution already ended. 
Las Vegas algorithm therefore improves both effectiveness and efficiency of the leader updating schedule, 
standing as a candidate mechanism to be employed in real recommender systems.

\section{Conclusion}

Online news websites are nowadays overloaded with a variety of fresh information. The hard task for readers is how to find 
the news they are really interested in. Recommender systems are a possible answer to this problem. 
A significant yet less noticed aspect is that social relationship is extremely useful in delivering news to 
potentially interested users. 

In this work we generalized a recently proposed news recommendation method~\cite{medo} to make it more efficient and effective. 
This method combines memory-based and social recommendation: recommendation scores of news items are computed using users' past ratings, 
though recommendations come only from the information sources a user selects. 
Another important feature of this model is that network's topology can be continually refined according to similarity 
of user tastes, in order for news to be efficiently delivered to the users that are most interested in them. 
We proposed new stochastic algorithms which are computationally cheap (and hence scalable with system's size) 
and provide an effective way to speed up the leader updating process. The next step will be to develop the described system on a real platform.

There are still some open questions to be answered. For instance, users do not differ only in their particular tastes but in many other aspects, 
like activity, reputation, generosity, and so on. News are also different in many aspects, like quality and freshness. 
Taking into account users' and news' heterogeneity may bring to a more realistic scenario. 
Another important point is that any real recommender system must face potential attacks 
from malicious users or communities, that may systematically introduce spam news or try to intentionally mislead the system. 
Measures of trustworthiness and reputation of users can be introduced to cope with these threats. 
\newline

We acknowledge helpful discussions with Chiho Yeung, Cihang Jin and Matus Medo. This work was partially supported by 
Swiss National Science Foundation (grant no. 200020-121848) and by the Future and Emerging Technologies programmes 
of the European Commission FP7-ICT-2007 (project LiquidPublication, grant no. 213360) and FP7-COSI-ICT (project QLectives, 
grant no. 231200). T.Z. acknowledges the National Natural Science Foundation of China under Grant Nos. 60973069.

\bibliographystyle{elsarticle-num}

\end{document}